# On Approximating Optimal Weighted Lobbying, and Frequency of Correctness versus Average-Case Polynomial Time[*]

Gábor Erdélyi[†]   Lane A. Hemaspaandra[‡]   Jörg Rothe[†]   Holger Spakowski[†]

March 18, 2007

**Abstract**

We investigate issues related to two hard problems related to voting, the optimal weighted lobbying problem and the winner problem for Dodgson elections. Regarding the former, Christian et al. [CFRS06] showed that optimal lobbying is intractable in the sense of parameterized complexity. We provide an efficient greedy algorithm that achieves a logarithmic approximation ratio for this problem and even for a more general variant—optimal weighted lobbying. We prove that essentially no better approximation ratio than ours can be proven for this greedy algorithm.

The problem of determining Dodgson winners is known to be complete for parallel access to NP [HHR97]. Homan and Hemaspaandra [HH06] proposed an efficient greedy heuristic for finding Dodgson winners with a guaranteed frequency of success, and their heuristic is a "frequently self-knowingly correct algorithm." We prove that every distributional problem solvable in polynomial time on the average with respect to the uniform distribution has a frequently self-knowingly correct polynomial-time algorithm. Furthermore, we study some features of probability weight of correctness with respect to Procaccia and Rosenschein's junta distributions [PR07].

[*]Supported in part by DFG grants RO 1202/9-1 and RO 1202/9-3, NSF grant CCF-0426761, the Alexander von Humboldt Foundation's TransCoop program, and a Friedrich Wilhelm Bessel Research Award. Work done in part while the second author was visiting Heinrich-Heine-Universität Düsseldorf. Some of the results of Section 3 of this paper were presented at the *First International Workshop on Computational Social Choice*, December 2006. This paper also appears as Univ. of Rochester Comp. Sci. Dept. Technical Report TR-2007-914.

[†]Institut für Informatik, Heinrich-Heine-Universität Düsseldorf, 40225 Düsseldorf, Germany. URLs: ccc.cs.uni-duesseldorf.de/{∼ erdelyi,∼ rothe,∼ spakowski}.

[‡]Department of Computer Science, University of Rochester, Rochester, NY 14627, USA. URL: www.cs.rochester.edu/u/lane.

# 1 Introduction

Preference aggregation and election systems have been studied for centuries in social choice theory, political science, and economics, see, e.g., Black [Bla58] and McLean and Urken [MU95]. Recently, these topics have become the focus of attention in various areas of computer science as well, such as artificial intelligence (especially with regard to distributed AI in multiagent settings), systems (e.g., for spam filtering), and computational complexity. Faliszewski et al. [FHHR] provides a survey of some recent progress in complexity-related aspects of elections.

This paper's topic is motivated by two hard problems that both are related to voting, the optimal weighted lobbying problem and the winner problem for Dodgson elections. Regarding the former problem, Christian et al. [CFRS06] defined its unweighted variant as follows: Given a 0-1 matrix that represents the No/Yes votes for multiple referenda in the context of direct democracy, a positive integer $k$, and a target vector (of the outcome of the referenda) of an external actor ("The Lobby"), is it possible for The Lobby to reach its target by changing the votes of at most $k$ voters? They proved the optimal lobbying problem complete for the complexity class W[2], thus providing strong evidence that it is intractable even for small values of the parameter $k$. However, The Lobby might still try to find an approximate solution efficiently. We propose an efficient greedy algorithm that establishes the first approximation result for the weighted version of this problem in which each voter has a price for changing his or her 0-1 vector to The Lobby's specification. Our approximation result applies to Christian et al.'s original optimal lobbying problem (in which each voter has unit price), and also provides the first approximation result for that problem. In particular, we achieve logarithmic approximation ratios for both these problems.

The Dodgson winner problem was shown NP-hard by Bartholdi, Tovey, and Trick [BTT89]. Hemaspaandra, Hemaspaandra, and Rothe [HHR97] optimally improved this result by showing that the Dodgson winner problem is complete for $P_{\parallel}^{NP}$, the class of problems solvable via parallel access to NP. Since these hardness results are in the worst-case complexity model, it is natural to wonder if one at least can find a heuristic algorithm solving the problem efficiently for "most of the inputs occurring in practice." Homan and Hemaspaandra [HH06] proposed a heuristic, called Greedy-Winner, for finding Dodgson winners. They proved that if the number of voters greatly exceeds the number of candidates (which in many real-world cases is a very plausible assumption), then their heuristic is a *frequently self-knowingly correct algorithm*, a notion they introduced to formally capture a strong notion of the property of "guaranteed success frequency" [HH06]. We study this notion in relation with average-case complexity. We also investigate Procaccia and Rosenschein's notion of deterministic heuristic polynomial time for their so-called junta distributions, a notion they introduced in their study of the "average-case complexity of manipulating elections" [PR07]. We show that under the junta definition, when stripped to its basic three properties, every NP-hard set is $\leq_m^p$-reducible to a set in deterministic heuristic polynomial time. We also show a very broad class of sets (including many NP-complete sets) to be in deterministic heuristic polynomial time. In an extended digression, we argue that the "average-case complexity" results of [PR07] are in fact not average-case complexity results, but rather are frequency of correctness—or, to be more precise, probability weight of correctness—results (as are also the results of Homan and Hemaspaandra).

This paper is organized as follows. In Section 2, we propose and analyze an efficient greedy algorithm for approximating the optimal weighted lobbying problem. In Section 3, we show that every problem solvable in average-case polynomial time with respect to the uniform distribution has a frequently self-knowingly correct polynomial-time algorithm, and we study Procaccia and Rosenschein's junta distributions. The appendix presents the heuristic Greedy-Score on which Greedy-Winner is based and the notion of frequently self-knowingly correct algorithm [HH06] as well as some needed technical definitions from average-case complexity theory [Lev86, Imp95, Gol97, Wan97].

## 2 Approximating Optimal Weighted Lobbying

### 2.1 Optimal Lobbying and its Weighted Version

Christian et al. [CFRS06] introduced and studied the following problem. Suppose there are $m$ voters who vote on $n$ referenda, and there is an external actor, which is referred to as "The Lobby" and seeks to influence the outcome of these referenda by making voters change their votes. It is assumed that The Lobby has complete information about the voters' original votes, and that The Lobby's budget allows for influencing the votes of a certain number, say $k$, of voters. Formally, the Optimal-Lobbying problem is defined as follows: Given an $m \times n$ 0-1 matrix $V$ (whose rows represent the voters, whose columns represent the referenda, and whose 0-1 entries represent No/Yes votes), a positive integer $k \leq m$, and a target vector $x \in \{0,1\}^n$, is there a choice of $k$ rows in $V$ such that by changing the entries of these rows the resulting matrix has the property that, for each $j$, $1 \leq j \leq n$, the $j$th column has a strict majority of ones (respectively, zeros) if and only if the $j$th entry of the target vector $x$ of The Lobby is one (respectively, zero) [CFRS06]?

Christian et al. [CFRS06] showed that Optimal-Lobbying (with respect to parameter $k$, the number of voters influenced by The Lobby) is complete for the complexity class W[2]; see, e.g., Downey and Fellows [DF99] and Flum and Grohe [FG06] for background on the theory of parameterized complexity and in particular for the definition of W[2].

This result is considered strong evidence that Optimal-Lobbying is intractable, even for small values of the parameter $k$. However, even though the optimal goal of The Lobby cannot be achieved efficiently, it might be approximable within some factor. That is, given an $m \times n$ 0-1 matrix $V$ and a target vector $x \in \{0,1\}^n$, The Lobby might try to reach its target by changing the votes of as few voters as possible.

We consider the more general problem Optimal-Weighted-Lobbying, where we assume that influencing the 0-1 vector of each voter $v_i$ exacts some price, $price(v_i) \in \mathbb{Q}$, where $\mathbb{Q}$ denotes the set of nonnegative rational numbers. In this scenario, The Lobby seeks to minimize the amount of money spent to reach its goal. The problem Optimal-Lobbying (redefined as an optimization problem rather than a parameterized problem) is the unit-prices special case of Optimal-Weighted-Lobbying, i.e., where $price(v_i) = 1$ for each voter $v_i$. It follows that Optimal-Weighted-Lobbying (redefined as a parameterized rather than an optimization problem, where the parameter is The Lobby's budget of money to be spent) inherits the W[2]-hardness lower bound from its special case Optimal-Lobbying, and that the logarithmic approximation algorithm we build for Optimal-Weighted-Lobbying will provide the same approximation ratio for Optimal-Lobbying.

In the remainder of this section, we describe and analyze an efficient greedy algorithm for approximating Optimal-Weighted-Lobbying.

## 2.2 A Greedy Algorithm for Optimal Weighted Lobbying

Let a matrix $V \in \{0,1\}^{m \times n}$ be given, where the columns $r_1, r_2, \ldots, r_n$ of $V$ represent the referenda and the rows $v_1, v_2, \ldots, v_m$ of $V$ represent the voters. Without loss of generality, we may assume that The Lobby's target vector is of the form $x = 1^n$ (and thus may be dropped from the problem instance), since if there is a zero in $x$ at position $j$, we can simply flip this zero to one and also flip the corresponding zeros and ones in column $r_j$.

For each column $r_j$, define the *deficit* $d_j$ to be the minimum number of zeros that need to be flipped to ones such that there are strictly more ones than zeros in this column. Let $D_0 = \sum_{j=1}^{n} d_j$ be the sum of all initial deficits.

Figure 1 gives the greedy algorithm, which proceeds by iteratively choosing a most "cost-effective" row of $V$ and flipping to ones all those zeros in this row that belong to columns with a positive deficit, until the deficits in all columns have decreased to zero. We assume that ties between rows with equally good cost-effectiveness are broken in any simple way, e.g., in favor of the tied $v_i$ with lowest $i$.

Let $R$ be the set of columns of $V$ whose deficits have already vanished at the beginning of an iteration, i.e., all columns in $R$ already have a strict majority of ones. Let $v_{i \upharpoonright R^c}$ denote the entries of $v_i$ restricted to those columns not in $R$, and let $\#_0(v_{i \upharpoonright R^c})$ denote the number of zeros in $v_{i \upharpoonright R^c}$. (For $i$ such that $\#_0(v_{i \upharpoonright R^c}) = 0$, we consider $price(v_i)/\#_0(v_{i \upharpoonright R^c})$ to be $+\infty$.) During an iteration, the *cost per flipped entry in row* $v_i$ (for decreasing the deficits in new columns by flipping $v_i$'s zeros to ones) is $price(v_i)/\#_0(v_{i \upharpoonright R^c})$. We say a voter $v_i$ is *more cost-effective* than a voter $v_j$ if $v_i$'s cost per flipped entry is less than $v_j$'s. When our algorithm chooses to alter a row $v_i$, we will think of its price being distributed equally among the new columns with decreased deficit, and at that instant will permanently associate with every flipped entry, $e_k$, in that row its portion of the cost, i.e., $cost(e_k) = price(v_i)/\#_0(v_{i \upharpoonright R^c})$.

Clearly, the greedy algorithm in Figure 1 always stops, and its running time is polynomial, since the while loop requires only linear (in the input size) time and has to be executed at most $D_0 = \sum_{j=1}^{n} d_j \leq n \cdot \lceil (m+1)/2 \rceil$ times (note that at most $\lceil (m+1)/2 \rceil$ flips are needed in each column to achieve victory for The Lobby's position).

Now, enumerate the $D_0$ entries of $V$ that have been flipped in the order in which they were flipped by the algorithm. Let $e_1, e_2, \ldots, e_{D_0}$ be the resulting enumeration. Let OPT be the money that would be spent by The Lobby for an optimal choice of voters such that its target is reached.

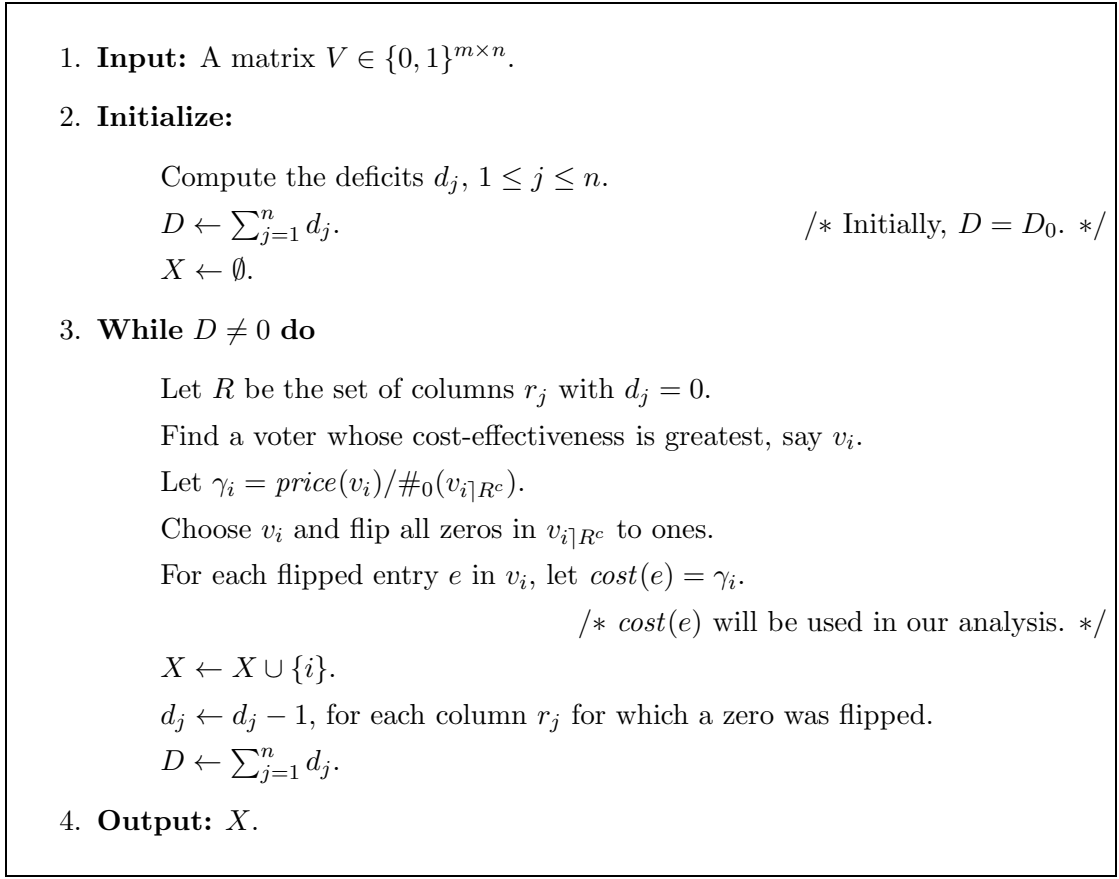

1. **Input:** A matrix $V \in \{0,1\}^{m \times n}$.

2. **Initialize:**

   Compute the deficits $d_j$, $1 \leq j \leq n$.
   $D \leftarrow \sum_{j=1}^{n} d_j$.  /* Initially, $D = D_0$. */
   $X \leftarrow \emptyset$.

3. **While $D \neq 0$ do**

   Let $R$ be the set of columns $r_j$ with $d_j = 0$.
   Find a voter whose cost-effectiveness is greatest, say $v_i$.
   Let $\gamma_i = price(v_i)/\#_0(v_{i \upharpoonright R^c})$.
   Choose $v_i$ and flip all zeros in $v_{i \upharpoonright R^c}$ to ones.
   For each flipped entry $e$ in $v_i$, let $cost(e) = \gamma_i$.
   /* $cost(e)$ will be used in our analysis. */
   $X \leftarrow X \cup \{i\}$.
   $d_j \leftarrow d_j - 1$, for each column $r_j$ for which a zero was flipped.
   $D \leftarrow \sum_{j=1}^{n} d_j$.

4. **Output:** $X$.

Figure 1: Greedy algorithm for Optimal-Weighted-Lobbying

**Lemma 2.1** *For each $k \in \{1, 2, \ldots, D_0\}$, we have $cost(e_k) \leq \mathrm{OPT}/(D_0 - k + 1)$.*

**Proof.** Let $I$ denote a voter set that realizes the optimal expenditure, OPT, for reducing the deficit to zero. Now, our analysis will follow the structure of the while loop of the algorithm. So consider the algorithm at some point where the current deficit, $D$, is strictly greater than zero and we are starting a pass through the while loop. So the entry we will next flip will be named $e_{D_0 - D + 1}$.

If we were to at this point consider changing all the rows associated with $I$ to all ones, this certainly would reduce the deficit to zero in the current matrix, as it in fact would even reduce the deficit to zero in the original matrix, and any prior passes through the while loop never flipped any entry in a way that went against the Lobby's goal (increased any deficit). Now, the first important thing to note is that there must be a collection $A$ of exactly $D$ zeros in the rows associated with $I$ such that flipping just those zeros reduces the deficit in the current matrix to zero. (This is clearly true due to the way deficits are computed and the separateness of the columns and their deficits.)

So by buying the rows of $I$ at cost OPT we certainly can flip all the $D$ entries composing $A$, i.e., we could image the cost as being distributed equally, and so we could view each such flipped entry as being purchased at cost $\text{OPT}/D$. However, the second important thing to note is that this means there is some element $i$ of $I$ that contains at least one element of $A$ such that for that element, at this moment, $price(v_i)/\#_0(v_i\rceil_{R^c})$ is at most $\text{OPT}/D$.[1] Since our algorithm chooses the most cost-effective row, it will choose a row with at least this cost-effectiveness.

Thus the first element of this iteration through the while loop, which will be $e_{D_0-D+1}$, is bought at cost at most $\text{OPT}/D$. So, for $k = D_0 - D + 1$ the claim of this lemma is satisfied, since $D_0 - (D_0 - D + 1) + 1 = D$.

However, note that each additional entry that we flip during this same pass through the while loop (e.g., some possibly empty prefix of $e_{D_0-D+2}$, $e_{D_0-D+3}$, etc.) will not only satisfy the claim of this lemma, but also will do even better, as it is (reducing the deficit by one and is) being bought at the cost of $\text{OPT}/D$, and the claim of the lemma was merely requiring that these additional elements be bought for, respectively, the strictly higher costs $\text{OPT}/(D-1)$, $\text{OPT}/(D-2)$, etc.

So, for each pass through the while loop, each entry flipped meets or beats the cost bound stated in this lemma. ❑

**Theorem 2.2** *The greedy algorithm presented in Figure 1 approximates the problem Optimal-Weighted-Lobbying with approximation ratio at most*

$$\sum_{i=1}^{D_0} \frac{1}{i} \leq 1 + \ln D_0 \leq 1 + \ln\left(n \left\lceil \frac{m+1}{2} \right\rceil\right).$$

**Proof.** The total price of the set of voters $X$ picked by the greedy algorithm is the sum of the costs of those entries flipped. That is,

$$price(X) = \sum_{i \in X} price(v_i) = \sum_{k=1}^{D_0} cost(e_k) \leq \left(1 + \frac{1}{2} + \cdots + \frac{1}{D_0}\right) \cdot \text{OPT},$$

where the last inequality follows from Lemma 2.1. ❑

---

[1]The reason is as follows. Consider for the moment associating with each element of each row of $I$ that contains at least one element of $A$ the portion of OPT indicated by the row's price divided by the number of elements of $A$ in that row. Clearly, the average of those $D$ element costs trivially must equal $\text{OPT}/D$, the overall average cost. So since the weighted average equals $\text{OPT}/D$, then at least one of the values being averaged must be less than or equal to $\text{OPT}/D$—and let us suppose that value is associated with row $i' \in I$. Since $A$ is a set of $D$ items that (starting from the current matrix) reduces the deficit by $D$, every element of $A$ in row $i'$ must reduce the deficit associated with its column by 1. And so every member of $A$ in row $i'$ must be a member of $v_{i'\rceil_{R^c}}$. So the value associated with each member of $A$ in row $i'$, which we already argued is at most $\text{OPT}/D$, is greater than or equal to the value $price(v_{i'})/\#_0(v_{i'\rceil_{R^c}})$ that the algorithm computes for each element of $A$—and indeed each element of $v_{i'\rceil_{R^c}}$—in that row.

|         | $r_1$ | $r_2$ | $r_3$ | $\cdots$ | $r_n$ | $price(v_i)$ |
|---------|-------|-------|-------|----------|-------|--------------|
| $v_1$   | 0     | 1     | 1     | $\cdots$ | 1     | 1            |
| $v_2$   | 1     | 0     | 1     | $\cdots$ | 1     | 1/2          |
| $v_3$   | 1     | 1     | 0     | $\cdots$ | 1     | 1/3          |
| $\vdots$ | $\vdots$ | $\vdots$ | $\vdots$ | $\ddots$ | $\vdots$ | $\vdots$ |
| $v_n$   | 1     | 1     | 1     | $\cdots$ | 0     | 1/n          |
| $v_{n+1}$ | 0   | 0     | 0     | $\cdots$ | 0     | $1+\epsilon$ |
| $v_{n+2}$ | 1   | 0     | 0     | $\cdots$ | 0     | 2            |
| $v_{n+3}$ | 0   | 1     | 0     | $\cdots$ | 0     | 2            |
| $v_{n+4}$ | 0   | 0     | 1     | $\cdots$ | 0     | 2            |
| $\vdots$ | $\vdots$ | $\vdots$ | $\vdots$ | $\ddots$ | $\vdots$ | $\vdots$ |
| $v_{2n+1}$ | 0  | 0     | 0     | $\cdots$ | 1     | 2            |

Table 1: A tight example for the greedy algorithm in Figure 1

Since the input size is lower-bounded by $m \cdot n$, Theorem 2.2 establishes a logarithmic approximation ratio for Optimal-Weighted-Lobbying (and also for Optimal-Lobbying). Note that the proof of Theorem 2.2 establishes an approximation ratio bound that is (sometimes nonstrictly) stronger than $\sum_{i=1}^{D_0} 1/i$. In particular, if the number of zeros flipped in successive iterations of the algorithm's while loop are $\ell_1, \ell_2, \ldots, \ell_p$, where $\ell_1 + \ell_2 + \cdots + \ell_p = D_0$, then the proof gives a bound on the approximation ratio of

$$\frac{\ell_1}{D_0} + \frac{\ell_2}{D_0 - \ell_1} + \frac{\ell_3}{D_0 - (\ell_1 + \ell_2)} + \cdots + \frac{\ell_p}{D_0 - (\ell_1 + \cdots + \ell_{p-1})} \;=\; \sum_{j=1}^{p} \frac{\ell_j}{D_0 - \sum_{k=1}^{j-1} \ell_k}.$$

This is strictly better than $\sum_{i=1}^{D_0} 1/i$ except in the case that each $\ell_j$ equals 1. And this explains why, in the example we are about to give that shows that the algorithm can at times yield a result with ratio essentially no better than $\sum_{i=1}^{D_0} 1/i$, each $\ell_j$ will equal 1.

Now, we show that the $\sum_{i=1}^{D_0} 1/i$ approximation ratio stated in Theorem 2.2 is essentially the best possible that can be stated for the greedy algorithm of Figure 1. Consider the example given in Table 1. The prices for changing the voters' 0-1 vectors are shown in the right-most column of Table 1: Set $price(v_i) = 1/i$ for each $i \in \{1, 2, \ldots, n\}$, set $price(v_i) = 2$ for each $i \in \{n+2, n+3, \ldots, 2n+1\}$, and set $price(v_{n+1}) = 1 + \epsilon$, where $\epsilon > 0$ is a fixed constant that can be set arbitrarily small. Note that, for each $j$, $1 \leq j \leq n$, we have $d_j = 1$, and hence $D_0 = n$.

When run on this input, our greedy algorithm sequentially flips, for $i = n, n-1, \ldots, 1$, the single zero-entry of voter $v_i$ to a one. Thus the total money spent is $1 + 1/2 + \cdots + 1/n = 1 + 1/2 + \cdots + 1/D_0$. On the other hand, the optimal choice consists of influencing just voter $v_{n+1}$ by flipping all of $v_{n+1}$'s entries to ones, which costs only $1 + \epsilon$.

# 3 Frequency of Correctness versus Average-Case Polynomial Time

## 3.1 A Motivation: How to Find Dodgson Winners Frequently

A Condorcet winner of an election is a candidate $i$ such that for each candidate $j \neq i$, a strict majority of the voters prefer $i$ to $j$. Not all elections have a Condorcet winner, but when a Condorcet winner exists, he or she is unique. In 1876, Dodgson [Dod76] proposed an election system that is based on a combinatorial optimization problem: An election is won by those candidates who are "closest" to being a Condorcet winner. More precisely, given a Dodgson election $(C, V)$,[2] every candidate $c$ in $C$ is assigned a score, denoted by DodgsonScore$(C, V, c)$, which gives the smallest number of sequential exchanges of adjacent preferences in the voters' preference orders needed to make $c$ a Condorcet winner with respect to the resulting preference orders. Whoever has the lowest Dodgson score wins.

The problem Dodgson-Winner is defined as follows: Given an election $(C, V)$ and a designated candidate $c$ in $C$, is $c$ a Dodgson winner in $(C, V)$? (The search version of this decision problem can easily be stated.) As mentioned earlier, Hemaspaandra et al. [HHR97] have shown that this problem is $P_{\|}^{NP}$-complete.

It certainly is not desirable to have an election system whose winner problem is hard, as only systems that can be evaluated efficiently are actually used in practice. Fortunately, there are a number of positive results on Dodgson elections and related systems as well. In particular, Bartholdi, Tovey, and Trick [BTT89] proved that for elections with a bounded number of candidates or voters Dodgson winners are asymptotically easy to determine. Fishburn [Fis77] proposed a "homogeneous" variant of Dodgson elections that Rothe, Spakowski, and Vogel [RSV03] proved to have a polynomial-time winner problem. McCabe-Dansted, Pritchard, and Slinko [MPS06] proposed a scheme (called Dodgson Quick) that approximates Dodgson's rule with an exponentially fast convergence. Homan and Hemaspaandra [HH06] proposed a greedy heuristic that finds Dodgson winners with a "guaranteed high frequency of success." To capture a strengthened version of this property formally, they introduced the notion of a "frequently self-knowingly correct algorithm" (see Appendix A for the formal definition and for their heuristic Greedy-Score).

## 3.2 On AvgP and Frequently Self-Knowingly Correct Algorithms

Our main result in this section relates polynomial-time benign algorithm schemes (see Definition B.2 in Appendix B) to frequently self-knowingly correct algorithms (see Definition A.1 in Appendix A). We show that every distributional problem that has a polynomial-time benign algorithm with respect to the uniform distribution scheme must also have a frequently self-knowingly correct polynomial-time algorithm. It follows that all uniformly distributed AvgP problems have a frequently self-knowingly correct polynomial-time algorithm.

---

[2] An election $(C, V)$ is given by a set $C$ of candidates and a set $V$ of voters, where each vote is specified by a preference order on all candidates and the underlying preference relation is strict (i.e., irreflexive and antisymmetric), transitive, and complete.

**Theorem 3.1** *Suppose that $\mathcal{A}(x, \delta)$ is a polynomial-time benign algorithm scheme for a distributional problem $f$ on $\hat{\mu}$ (the standard uniform distribution, see Appendix B). Then there is a frequently self-knowingly correct polynomial-time algorithm $\mathcal{A}'$ for $f$.*

**Proof.** Let $\delta(n) = 1/(n+1)^3$. Define algorithm $\mathcal{A}'$ as follows:

1. On input $x \in \Sigma^*$, simulate $\mathcal{A}(x, \delta(|x|))$.

2. If $\mathcal{A}(x, \delta(|x|))$ outputs ?, then output (*anything*, *"maybe"*).

3. If $\mathcal{A}(x, \delta(|x|))$ outputs $y \in T$, where $y \neq ?$, then output $(y, \text{``definitely''})$.

By Definition B.2 (which is given in Appendix B), algorithm $\mathcal{A}'$ runs in polynomial time. It remains to show that $\mathcal{A}'$ is frequently self-knowingly correct.

Fix an arbitrary $n \in \mathbb{N}^+$. Now, we must be careful regarding the fact that Impagliazzo's definition of benign algorithm schemes and its "$\delta$" guarantees are all with regard to drawing not over inputs of a given length (which is what we wish to consider) but rather regarding drawing from inputs *up to and including* a given length. Thus, there is some danger that even if a benign algorithm performs well when its length parameter is $n$ (meaning related to strings of length up to and including $n$), that such a "good" error frequency might be due not to goodness at length $n$ but rather to goodness at lengths $n-1$, $n-2$, and so on. However, if one looks carefully at the relative weights of the different lengths this is at most a quadratically weighted effect (that is, the distribution's probability weight at length $n$ is just quadratically less than the weight summed over all lengths less than $n$), and so our choice of $\delta(n) = 1/(n+1)^3$ is enough to overcome this.

Let us now handle that rigorously. Recall that $n$ is fixed and arbitrary. Let us set the constant (for fixed $n$) $\delta'$ to be $1/(n+1)^3$. So, clearly

$$\mathrm{Prob}_{\hat{\mu}_{\leq n}}[\mathcal{A}(x, \delta') = ?] =$$

$$\frac{\sum_{i=1}^{n-1} 1/(i(i+1))}{\sum_{i=1}^{n} 1/(i(i+1))} \mathrm{Prob}_{\hat{\mu}_{\leq n-1}}[\mathcal{A}(x, \delta') = ?] + \frac{1/(n(n+1))}{\sum_{i=1}^{n} 1/(i(i+1))} \mathrm{Prob}_{\hat{\mu}_n}[\mathcal{A}(x, \delta') = ?].$$

Since $\mathcal{A}$ is a benign algorithm scheme, $\mathrm{Prob}_{\hat{\mu}_{\leq n}}[\mathcal{A}(x, \delta') = ?] \leq \delta'$. So, combining this and the above equality, and solving for $\mathrm{Prob}_{\hat{\mu}_n}[\mathcal{A}(x, \delta') = ?]$, we have

$$\mathrm{Prob}_{\hat{\mu}_n}[\mathcal{A}(x, \delta') = ?] \leq$$

$$\frac{\sum_{i=1}^{n} 1/(i(i+1))}{1/(n(n+1))} \left( \delta' - \frac{\sum_{i=1}^{n-1} 1/(i(i+1))}{\sum_{i=1}^{n} 1/(i(i+1))} \mathrm{Prob}_{\hat{\mu}_{\leq n-1}}[\mathcal{A}(x, \delta') = ?] \right).$$

And so, clearly, $\mathrm{Prob}_{\hat{\mu}_n}[\mathcal{A}(x, \delta') = ?] \leq n(n+1)\delta' = n(n+1)/(n+1)^3$. So

$$\lim_{n \to \infty} \frac{\|\{x \in \Sigma^n \mid \mathcal{A}'(x) \in T \times \{\text{``maybe''}\}\}\|}{\|\Sigma^n\|} = 0,$$

which completes the proof. ❑

**Corollary 3.2** *Every distributional problem that under the standard uniform distribution is in* AvgP *has a frequently self-knowingly correct polynomial-time algorithm.*

**Proof.** Impagliazzo proved that any distributional problem on input ensemble $\mu_n$ is in AvgP if and only if it has a polynomial-time benign algorithm scheme; see Proposition 2 in [Imp95]. The claim now follows from Theorem 3.1. ❑

It is easy to see that the converse implication of that in Corollary 3.2 is not true.

**Proposition 3.3** *There exist (distributional) problems with a frequently self-knowingly correct polynomial-time algorithm that are not in* AvgP *under the standard uniform distribution.*

**Proof.** For instance, one can define a problem that consists only of strings in $\{0\}^*$ encoding the halting problem. This problem is clearly not in AvgP, yet it is frequently self-knowingly correct. ❑

### 3.3 A Basic Junta Distribution for SAT (and a Digression on Whether Heuristic Polynomial-Time Algorithms Yield Good Average-Case Complexity)

Procaccia and Rosenschein [PR07] introduced "junta distributions" in their study of NP-hard manipulation problems for elections. The goal of a junta is to be such a hard distribution (that is, to focus so much weight on hard instances) that, loosely put, if a problem is easy relative to a junta then it will be easy relative to any reasonable distribution (such as the uniform distribution). This is a goal, not (currently) a theorem; Procaccia and Rosenschein [PR07] do not formally establish this, but rather seek to give a junta definition that might satisfy this. Their paper in effect encourages others to weigh in and study the suitability of the notion of a junta and the notion built on top of it, heuristic polynomial time. Furthermore, they repeatedly state that their theory is one of average-case complexity.

Regarding the latter claim, we in Footnote 4 point out that it is inaccurate to describe their theory as one of average-case complexity. Their theory adds to the study of frequency of correctness the notion of probability weight of correctness. This is a very valuable direction, but we point out that it is neither explicitly about, nor does it seem to implicitly yield claims about, average-case complexity. Their paper states that work of Conitzer and Sandholm [CS06] is also about average-case complexity but, similarly, we mention that that work is not about average-case complexity; it is about (and carefully and correctly frames itself as being about) frequency of correctness. We do not mean this as a weakness: We feel that frequency of (or probability weight of) correctness, most especially when as in the work of Homan and Hemaspaandra [HH06] the algorithm is "self-knowingly" correct a guaranteed large portion of the time, is an interesting and important direction.

Regarding Procaccia and Rosenschein's notion of juntas, they state three "basic" conditions for a junta, and then give two additional ones that are tailored specifically to the needs of NP-hard voting manipulation problems. They state their hope that their scheme will extend more generally, using the three basic conditions and potentially additional conditions, to other mechanism problems. One might naturally wonder whether their junta/heuristic polynomial-time approach applies more generally to studying the probability weight of correctness for NP-hard problems, since their framework in effect (aside from the two "additional" junta conditions just about voting manipulation) is a general one relating problems to probability weight of correctness. We first carefully note that in asking this we are taking their notion beyond the realm for which it was explicitly designed, and so we do not claim to be refuting any claim of their paper. What we will do, however, is show that the three basic conditions for a junta are sufficiently weak that one can construct a junta relative to which the standard NP-complete problem SAT—and a similar attack can be carried out on a wide range of natural NP-complete problems—has a deterministic heuristic polynomial-time algorithm. So if one had faith in the analog of their approach, as applied to SAT, one would have to believe that under essentially every natural distribution SAT is easy (in the sense that there is an algorithm with a high probability weight of correctness under that distribution). Since the latter is not widely believed, we suggest that the right conclusion to draw from the main result of this section is simply that if one were to hope to effectively use the notion of juntas and heuristic polynomial time on typical NP-complete sets, one would almost certainly have to go beyond the basic three conditions and add additional conditions. Again, we stress that Procaccia and Rosenschein didn't focus on examples this far afield, and even within the world of mechanisms implied that unspecified additional conditions beyond the core three might be needed when studying problems other than voting manipulation problems. This section's contribution is to give a construction indicating that the core three junta conditions, standing on their own, seem too weak.

Since we will use the Procaccia–Rosenschein junta notion in a more general setting than merely manipulation problems, we to avoid any chance of confusion will use the term "basic junta" to denote that we have removed the word "manipulation" and that we are using their three "basic" properties, and not the two additional properties that are specific to voting manipulation. Our definition of "deterministic heuristic polynomial-time algorithm" is identical to theirs, except we have replaced the word "junta" with "basic junta"—and so again we are allowing their notion to be extended beyond just manipulation and mechanism problems.

**Definition 3.4 (see [PR07])** *Let $\mu = \{\mu_n\}_{n \in \mathbb{N}}$ be a distribution over the possible instances of an NP-hard problem L. (In this model, each $\mu_n$ sums to 1 over all length n instances.[3]) We say $\mu$ is a* basic junta distribution *if and only if $\mu$ has the following properties:*

---

[3]We say this because in the Procaccia–Rosenschein work, they state that each $\mu_n$ is a distribution, all their work and notions are based on looking at a single length at a time, and in their example of building a junta they do nothing at all to address relative weights between different lengths (and so a global distribution, i.e., one over $\Sigma^*$, is not being defined). In addition to these three reasons, if one were to try to interpret the notion as saying that there is a (Levin-like) single distribution over all lengths, one would in their definition of junta have foundational problems when that single distribution put no weight on any strings of a given length, as one would be faced with conditioning on a set of probability weight zero, which is not well-defined.

1. **Hardness:** *The restriction of $L$ to $\mu$ is the problem whose possible instances are only $\bigcup_{n\in\mathbb{N}}\{x \mid |x| = n \text{ and } \mu_n(x) > 0\}$. Deciding this restricted problem is still* NP-*hard.*

2. **Balance:** *There exist constants $c > 1$ and $N \in \mathbb{N}$ such that for all $n \geq N$ and for all instances $x$, $|x| = n$, we have $1/c \leq \mathrm{Prob}_{\mu_n}[x \in L] \leq 1 - 1/c$.*

3. **Dichotomy:** *There exists some polynomial $p$ such that for all $n$ and for all instances $x$, $|x| = n$, either $\mu_n(x) \geq 2^{-p(n)}$ or $\mu_n(x) = 0$.*

*Let $(L, \mu)$ be a distributional decision problem (see Definition B.1 in Appendix B). An algorithm $\mathcal{A}$ is said to be a* deterministic heuristic polynomial-time algorithm *for $(L, \mu)$ if $\mathcal{A}$ is a deterministic polynomial-time algorithm and there exist a polynomial $q$ and $N \in \mathbb{N}$ such that for each $n \geq N$,*

$$\mathrm{Prob}_{\mu_n}[x \notin L \iff \mathcal{A} \text{ accepts } x] < \frac{1}{q(n)}.$$

We in the footnote to this sentence digress to suggest that this should not be viewed as providing an average-case complexity theory.[4]

---

On the other hand, Procaccia–Rosenschein do use the phrase "distributional problem," and we mention that in that notion, as it is typically used, the distribution is global; however, for the three reasons mentioned above, the Procaccia–Rosenschein work's use of it is most coherently interpreted as, when the words are used, the "distribution" being simply a collection of distributions, one per length. We mention in passing that our main theorem of this section, Theorem 3.6, remains true—though one has to shift the values in its proof a bit—even under the alternate interpretation of one global distribution. On the other hand, results such as nonclosure under polynomial-time isomorphisms potentially might not hold under that alternate model.

[4]Procaccia and Rosenschein [PR07] in the title and body of their paper repeatedly describe their theory as being an "average-case complexity" theory for manipulation. We note that this is inaccurate within any natural interpretation of the notion of "average-case complexity." We mean this not merely because their theory lacks what certainly must be a part of any average-case complexity theory, namely, *taking an average of complexities*. Indeed, their theory is actually an approach to adding probability weights to a frequency of correctness approach. In brief, they look at probability weight (relative to some distribution) of correctness. However, they require correctness only for $1 - 1/poly$ probability weight at each length. This is quite a lot of probability weight, but we now note that that seems not anywhere near enough to ensure good average-case complexity. Let us consider taking a deterministic heuristic polynomial-time algorithm for a problem and trying to build from it an algorithm for the problem (i.e., one that is correct on all instances) that has good average-case running time. First, notice that one is dead from the start, as deterministic heuristic polynomial-time algorithms, though very frequently correct, are not guaranteed to know any easily recognized broad set of inputs on which they are guaranteed to be correct. But for the sake of argument, let us suppose that for our given deterministic heuristic polynomial-time algorithm we by good luck have that there is a deterministic polynomial-time set, having at each length probability weight under the junta at least $1 - 1/poly$, such that for each element of this set the deterministic heuristic polynomial-time algorithm is correct. Of course, for the remaining $1/poly$ of the weight the algorithm might be correct or not correct—no guarantees. Even with this strong extra assumption (which is basically tossing in a Homan–Hemaspaandra-like self-knowing correctness property, see [HH06]), the average-case time analysis doesn't come out happily. Note that for the remaining $1/poly$ of the probability weight (and we are assuming that this is not just an upper bound but that one might actually have about this much weight for these bad cases), one would have to potentially brute force those, and as Procaccia and Rosenschein focus on NP-hard

We now explore their notion of deterministic heuristic polynomial time[5] and their notion of junta, both however viewed for general NP problems and using the "basic" three conditions. We will note that the notion in such a setting is in some senses not restrictive enough and in other senses is too restrictive. Let us start with the former. We need a definition.

**Definition 3.5** *We will say that a set $L$ is* well-pierced *(respectively,* uniquely well-pierced*) if there exist sets $Pos \in \text{P}$ and $Neg \in \text{P}$ such that $Pos \subseteq L$, $Neg \subseteq \overline{L}$, and there is some $N \in \mathbb{N}$ such that at each length $n \geq N$, each of Pos and Neg has at least one string at length $n$ (respectively, each of Pos and Neg has exactly one string at length $n$).*

Each uniquely well-pierced set is well-pierced. Note that, under quite natural encodings, such NP-complete sets as, for example, SAT certainly are well-pierced and uniquely well-

---

problems, each such brute-forcing would seem to potentially take exponential time. So, very loosely put, and under all the assumptions we are making (e.g., about having to use brute force and so on), the expected time (over their own distribution) one gets is roughly $\left(1 - \frac{1}{poly}\right) \cdot poly' + \frac{exponential}{poly}$. And, critically, that is exponential. (What we just argued is that, very informally, in the model of looking at junta-weighted average time over the strings of each length and looking at the asymptotics of that, the obvious attempt to convert a deterministic heuristic polynomial-time algorithm into an algorithm (i.e., a correct program for the problem) with good average-case running time yields an exponential average. However, it is true that such asymptotics of averages over each length have in other settings some undesirably properties, see, e.g., [Gol97]. Nonetheless, the $1/poly$ weight of the exponential time here is so bad that going into an even more Levin-like setting by tamping down on the runtimes by adding an "$\epsilon$" exponent still would not seem, if done naturally, to tame the exponential nature of the average time.)

So in summary Procaccia and Rosenschein do not build a theory of average-case complexity, but rather shift the nature of "frequency of correctness" approaches to focus instead on "probability weight of correctness (relative to some distribution)"—which is a quite natural shift to look at. We mention in passing that the papers of Homan and Hemaspaandra [HH06] and of Conitzer and Sandholm [CS06] are about frequency of correctness—and are quite explicit that that, and not average-case complexity, is what they are about. This ends our informal digression/discussion of frequency of correctness versus average-case complexity.

However, as a quick postscript just for those interested, we discuss this issue, very informally, regarding Homan and Hemaspaandra [HH06]. The reason we do so is that that paper seems to ensure not just an at most $1/poly$ weight of bad cases, but in fact an at most $1/exponential$ proportion (note: it is in a uniform-like model) of bad cases. And so one might hope that it might yield average-case polynomial asymptotics. However, this seems not to be the case. In more detail, yet speaking very informally: Note that if we choose uniformly a random $m$-candidate, $n$-voter Dodgson election instance $I$ and use the Greedy-Winner algorithm on it and in the case of a "maybe" then brute-force it, the expected running time will be, where $p$ is the polynomial run time of Greedy-Winner and $cm^n$ is our brute-force time for a brute-force solution, $\left(1 - 2(m^2 - m)e^{-\frac{n}{8m^2}}\right) \cdot \text{E}_{good(m,n)}[p(|I|)] + 2(m^2-m)e^{-\frac{n}{8m^2}} \cdot \text{E}_{unknown(m,n)}[p(|I|) + cm^n]$, where $\text{E}[\cdot]$ denotes expectation, $good(m,n)$ denotes the set of $m$-candidate, $n$-voter election instances for which the algorithm is self-knowingly correct, and $unknown(m,n)$ denotes the set of $m$-candidate, $n$-voter election instances for which the algorithm is not self-knowingly correct. Note that, assuming that the instances with given $m$ and $n$ are tightly clustered in the lengths they encode to (and such clustering will typically be the case if we don't allow flexible text-string names as part of the input), we have, basically since $e^{-\frac{n}{8m^2}} \cdot m^n = e^{(n \ln m) - \frac{n}{8m^2}}$, that the expected time is exponential (as $n$ grows).

[5]They credit their notion as being "inspired by Trevisan [Tre02] (there the same name is used for a somewhat different definition)." We mention in passing as an even earlier source for the same name, though also attached to a different definition than that of Procaccia and Rosenschein, Section 3 of [Imp95].

pierced. (All this says is that, except for a finite number of exceptional lengths, there is one special string at each length that can easily, uniformly be recognized as in the set and one that can easily, uniformly be recognized as not in the set.) Indeed, under quite natural encodings, undecidable problems such as the halting problem are uniquely well-pierced.

Recall that juntas are defined in relation to an infinite list of distributions, one per length (so $\mu = \{\mu_n\}_{n \in \mathbb{N}}$). The Procaccia and Rosenschein definition of junta does not explicitly put computability or uniformity requirements on such distributions in the definition of junta, but it is useful to be able to make claims about that. So let us say that such a distribution is *uniformly computable in polynomial time* (respectively, is *uniformly computable in exponential time*) if there is a polynomial-time function (respectively, an exponential-time function) $f$ such that for each $i$ and each $x$, $f(i, x)$ outputs the value of $\mu_i(x)$ (say, as a rational number—if a distribution takes on other values, it simply will not be able to satisfy our notion of good uniform time).

**Theorem 3.6** *Let $A$ be any NP-hard set that is well-pierced. Then there exists a basic junta distribution relative to which $A$ has a deterministic heuristic polynomial-time algorithm (indeed, it even has a deterministic heuristic polynomial-time algorithm whose error weight is bounded not merely by $1/\text{poly}$ as the definition requires, but is even bounded by $1/2^{n^2-n}$). Furthermore, the junta is uniformly computable in exponential time, and if we in addition assume that $A$ is uniquely well-pierced, the junta is uniformly computable in polynomial time.*

It follows that, under quite natural encodings, almost any natural set is in deterministic heuristic polynomial time. For example, SAT is and the halting problem is, both under natural encodings.[6] All it takes is for the given set to have at all but a finite number of lengths at least one element each that are uniformly easily recognizable as being in and out of the set.

**Proof.** Let $A$ be well-pierced. So there exists an $N$, and sets *Pos* and *Neg*, that satisfy the definition of well-pierced. For each $n \geq N$, let $Pos(n)$ denote the lexicographically smallest length $n$ string in *Pos* and let $Neg(n)$ denote the lexicographically smallest length $n$ string in *Neg*.

Define the distribution $\nu = \{\nu_n\}_{n \in \mathbb{N}}$ as follows:

1. For each length $n \geq N$, put weight $1/2^{n^2}$ on all length $n$ strings other than $Pos(n)$ and $Neg(n)$, and put weight $\frac{1}{2}\left(1 - \frac{2^n-2}{2^{n^2}}\right)$ on each of $Pos(n)$ and $Neg(n)$.

---

[6] Again, a potential problem when dealing with such claims is details of encoding. For example, if SAT is encoded in such a way that the vast majority of the strings (namely all but at most a $1/poly$ portion of the strings) of each length are obviously syntactically illegal (and such encodings can indeed be totally natural), then an astute reader might well ask, "Isn't any algorithm that accepts the empty set a deterministic heuristic polynomial-time algorithm for SAT, relative to the uniform distribution, which obviously is a basic junta." However, this reasoning is flawed. If that many strings of each length are obviously syntactically illegal and we are using the uniform distribution, then the balance condition for juntas is violated. So the balance condition blocks that argument, and indeed this type of blocking is *precisely* why Procaccia and Rosenschein [PR07] have the balance condition.

2. For each length $n < N$, let $\nu_n$ be the uniform distribution over that length, i.e., each length $n$ string has weight $1/2^n$.

We now show that $\nu$ is a basic junta distribution.

1. Hardness: Since $\bigcup_{n \in \mathbb{N}} \{x \mid |x| = n \text{ and } \nu_n(x) > 0\}$ equals $\Sigma^*$, the restriction of $A$ to $\nu$ equals $A$, and so is still NP-hard.

2. Balance: Since for each length $n \geq N$ both $Pos(n) \in A$ and $Neg(n) \notin A$ have almost half of the probability weight of all length $n$ strings (namely, each has $\frac{1}{2}\left(1 - \frac{2^n - 2}{2^{n^2}}\right)$), $\nu$ is balanced.

3. Dichotomy: Since for all $n \geq N$ and for all $x$, $|x| = n$, we have $\nu_n(x) \geq 2^{-n^2}$, and for all $n < N$ and for all $x$, $|x| = n$, we have $\nu_n(x) \geq 2^{-n}$, dichotomy is satisfied.

Note that the junta is uniformly computable in exponential time, and if $A$ is uniquely well-pierced then the junta is uniformly computable in polynomial time.

Our deterministic heuristic polynomial-time algorithm for $(A, \nu)$ works as follows: On inputs that are a $Pos(n)$, it accepts; on inputs that are a $Neg(n)$, it rejects; and on every other input, it (for specificity, though it does not matter) accepts.

For each $n \geq N$, the error probability of this algorithm on inputs of length $n$ is at most $(2^n - 2)/2^{n^2} \leq 1/2^{n^2 - n}$. ❏

In the proof we achieve the error bound $1/2^{n^2-n}$ stated in Theorem 3.6. However, this bound can easily be strengthened to $1/2^{n^k - n}$, for each fixed constant $k$, by altering the proof. Note that the altered algorithm will depend on $k$.

Our point is not that this construction is difficult. Rather, our point is that this construction indicates that the basic three junta conditions on their own can be short-circuited, and thus a stronger set of conditions would be needed to seek to create a Procaccia–Rosenschein-type program against, e.g., SAT. More generally, one should probably be exceedingly skeptical about any distribution or distribution type that is being proposed—without proof—as perhaps being so hard that it seeks to "convincingly represent all other distributions with respect to average-case analysis" [PR07, p. 163]. Again, we should stress that Procaccia and Rosenschein are clear that this is a hope not a claim, that they repeatedly stress that the approach may be controversial, and that their focus is on manipulation/mechanism issues.

Loosely put, the above result says that the basic junta conditions are in some ways overinclusive. We also note that the definition of junta, and the issue of when we will have a heuristic polynomial-time algorithm, are exceedingly sensitive to details of encoding.[7] We mention quickly two such effects, one that indirectly suggests overinclusiveness and one that suggests underinclusiveness.

---

[7]In contrast, the "$\epsilon$" exponent and $|x|$ denominator (see Definition B.1 in Appendix B) in Levin's [Lev86] theory of AvgP, average-case polynomial-time, were precisely designed, in that different setting, to avoid such problems—problems that one gets by following the type of asymptotic focus on one length at a time that the Procaccia and Rosenschein model adopts. On the other hand, even Levin's theory has many subtleties and downsides, and to this day has not found anything resembling the type of widespread applicability of NP-completeness theory; see any of the many surveys on that topic.

As to the former, note that *every* NP-hard set is $\leq_m^p$-reducible to a set that is in deterministic heuristic polynomial time. This applies even to undecidable NP-hard sets, such as $\text{SAT} \oplus \text{HP} =_{def} \{0x \mid x \in \text{SAT}\} \cup \{1y \mid y \in \text{HP}\}$, where HP denotes the halting problem. The proof is nearly immediate. Given an NP-hard set $A$ (over some alphabet $\Sigma$ that has cardinality at least two, and w.l.o.g. we assume that 0 and 1 are letters of $\Sigma$), note that $A \leq_m^p$-reduces to the set $A' = \{00x \mid x \in \Sigma^*\} \cup \{1x1^{|x|^2+2} \mid x \in A\}$, and that $A'$ is easily seen to be in deterministic heuristic polynomial time (indeed, with error bound not just $1/poly$ but even $1/exponential$), in particular via the junta (relative to $A'$) that is the uniform distribution.

Regarding underinclusiveness, note that under the definition of junta, no set that at an infinite number of lengths either has all strings or has no strings can have a deterministic heuristic polynomial-time algorithm, since for such sets the balance condition of the notion of a junta can never be satisfied. It follows easily that the notion of having a deterministic heuristic polynomial-time algorithm is not even closed under polynomial-time isomorphisms.[8]

## 4 Conclusions

Christian et al. [CFRS06] introduced the optimal lobbying problem and showed it complete for W[2], and so generally viewed as intractable in the sense of parameterized complexity. In Section 2, we proposed an efficient greedy algorithm for approximating the optimal solution of this problem, even if generalized by assigning prices to voters. This greedy algorithm achieves a logarithmic approximation ratio and we prove that that is essentially the best approximation ratio that can be proven for this algorithm. We mention as an interesting open issue whether more elaborate algorithms can achieve better approximation ratios.

Section 3 studied relationships between average-case polynomial time, benign algorithm schemes, and frequency (and probability weight) of correctness. We showed that all problems having benign algorithm schemes relative to the uniform distribution (and thus all sets in average-case polynomial time relative to the uniform distribution) have frequently self-knowingly correct algorithms. We also studied, when limited to the "basic" three junta conditions, the notion of junta distributions and of deterministic heuristic polynomial time, and we showed that they admit some extreme behaviors. We argued that deterministic heuristic polynomial time should not be viewed as a model of average-case complexity.

**Acknowledgments:** We are deeply grateful to Chris Homan for his interest in this work

---

[8]To be extremely concrete, the NP-complete set $B = \{00x \mid x \in \Sigma^*\} \cup \{1x1^{|x|^2+2} \mid x \in \text{SAT}\}$ is (as per the above) easily seen to be in deterministic heuristic polynomial time, but the NP-complete set $B' = \{xx \mid x \in \text{SAT}\}$, though it is by standard techniques polynomial-time isomorphic to $B$ (see [BH77]), is not in deterministic heuristic polynomial time. If the reader wonders why we did not simply use two P sets, the reason is, under to the Procaccia–Rosenschein definition, one needs NP-hardness to have a junta, and one needs a junta to put something in deterministic heuristic polynomial time.

and for many inspiring discussions on computational issues related to voting. We also thank the anonymous COMSOC 2006 workshop referees for their helpful comments.# References

[BH77]    L. Berman and J. Hartmanis. On isomorphisms and density of NP and other complete sets. *SIAM Journal on Computing*, 6(2):305–322, 1977.

[Bla58]   D. Black. *The Theory of Committees and Elections*. Cambridge University Press, 1958.

[BTT89]   J. Bartholdi III, C. Tovey, and M. Trick. Voting schemes for which it can be difficult to tell who won the election. *Social Choice and Welfare*, 6(2):157–165, 1989.

[CFRS06]  R. Christian, M. Fellows, F. Rosamond, and A. Slinko. On complexity of lobbying in multiple referenda. In U. Endriss and J. Lang, editors, *First International Workshop on Computational Social Choice (COMSOC 2006)*, pages 87–96 (workshop notes). Universiteit van Amsterdam, December 2006.

[CS06]    V. Conitzer and T. Sandholm. Nonexistence of voting rules that are usually hard to manipulate. In *Proceedings of the 21st National Conference on Artificial Intelligence*. AAAI Press, July 2006.

[DF99]    R. Downey and M. Fellows. *Parameterized Complexity*. Springer-Verlag, Berlin, Heidelberg, New York, 1999.

[Dod76]   C. Dodgson. A method of taking votes on more than two issues. Pamphlet printed by the Clarendon Press, Oxford, and headed "not yet published" (see the discussions in [MU95, Bla58], both of which reprint this paper), 1876.

[FG06]    J. Flum and M. Grohe. *Parameterized Complexity Theory*. EATCS Texts in Theoretical Computer Science. Springer-Verlag, Berlin, Heidelberg, 2006.

[FHHR]    P. Faliszewski, E. Hemaspaandra, L. Hemaspaandra, and J. Rothe. A richer understanding of the complexity of election systems. In S. Ravi and S. Shukla, editors, *Fundamental Problems in Computing: Essays in Honor of Professor Daniel J. Rosenkrantz*. Springer. To appear. Available as Technical Report cs.GT/0609112, ACM Computing Research Repository (CoRR), September 2006.

[Fis77]   P. Fishburn. Condorcet social choice functions. *SIAM Journal on Applied Mathematics*, 33(3):469–489, 1977.17

[Gol97]   O. Goldreich. Note on Levin's theory of average-case complexity. Technical Report TR97-058, Electronic Colloquium on Computational Complexity, November 1997.

[HH06]   C. Homan and L. Hemaspaandra. Guarantees for the success frequency of an algorithm for finding Dodgson-election winners. In *Proceedings of the 31st International Symposium on Mathematical Foundations of Computer Science*, pages 528–539. Springer-Verlag *Lecture Notes in Computer Science #4162*, August/September 2006.

[HHR97]   E. Hemaspaandra, L. Hemaspaandra, and J. Rothe. Exact analysis of Dodgson elections: Lewis Carroll's 1876 voting system is complete for parallel access to NP. *Journal of the ACM*, 44(6):806–825, November 1997.

[Imp95]   R. Impagliazzo. A personal view of average-case complexity. In *Proceedings of the 10th Structure in Complexity Theory Conference*, pages 134–147. IEEE Computer Society Press, 1995.

[Lev86]   L. Levin. Average case complete problems. *SIAM Journal on Computing*, 15(1):285–286, 1986.

[MPS06]   J. McCabe-Dansted, G. Pritchard, and A. Slinko. Approximability of Dodgson's rule. In U. Endriss and J. Lang, editors, *First International Workshop on Computational Social Choice (COMSOC 2006)*, pages 331–344 (workshop notes). Universiteit van Amsterdam, December 2006.

[MU95]   I. McLean and A. Urken. *Classics of Social Choice*. University of Michigan Press, Ann Arbor, Michigan, 1995.

[PR07]   A. Procaccia and J. Rosenschein. Junta distributions and the average-case complexity of manipulating elections. *Journal of Artificial Intelligence Research*, 28:157–181, 2007.

[RSV03]   J. Rothe, H. Spakowski, and J. Vogel. Exact complexity of the winner problem for Young elections. *Theory of Computing Systems*, 36(4):375–386, June 2003.

[Tre02]   L. Trevisan. Lecture notes on computational complexity. www.cs.berkeley.edu/~luca/notes/complexitynotes02.pdf (Lecture 12), 2002.

[Wan97]   J. Wang. Average-case computational complexity theory. In L. Hemaspaandra and A. Selman, editors, *Complexity Theory Retrospective II*, pages 295–328. Springer-Verlag, 1997.

# A  Homan and Hemaspaandra's Frequently Self-Knowingly Correct Greedy Algorithm

Homan and Hemaspaandra [HH06] proposed the following definition of a new type of algorithm to capture the notion of "guaranteed high success frequency" formally.

**Definition A.1 ([HH06])**  1. *Let $f : S \to T$ be a function, where $S$ and $T$ are sets. We say an algorithm $\mathcal{A} : S \to T \times \{$ "definitely", "maybe"$\}$ is* self-knowingly correct *for $f$ if, for each $s \in S$ and $t \in T$, whenever $\mathcal{A}$ on input $s$ outputs $(t,$ "definitely"$)$ then $f(s) = t$.*

2. *An algorithm $\mathcal{A}$ that is self-knowingly correct for $g : \Sigma^* \to T$ is said to be* frequently self-knowingly correct *for $g$ if*

$$\lim_{n \to \infty} \frac{\|\{x \in \Sigma^n \mid A(x) \in T \times \{\text{"maybe"}\}\}\|}{\|\Sigma^n\|} = 0.$$

In their paper [HH06], Homan and Hemaspaandra present two frequently self-knowingly correct polynomial-time algorithms, which they call Greedy-Score and Greedy-Winner. Since Greedy-Winner can easily be reduced to Greedy-Score, we focus on Greedy-Score only and briefly describe the intuition behind this algorithm; for full detail, we refer to [HH06]. (But both heuristics work well tremendously often—in a formal sense of the notion—provided that the number of voters greatly exceeds the number of candidates.)

If $(C, V)$ is an election and $c$ is some designated candidate in $C$, we call $(C, V, c)$ a *Dodgson triple*. Given a Dodgson triple $(C, V, c)$, Greedy-Score determines the Dodgson score of $c$ with respect to the given election $(C, V)$. We will see that there are Dodgson triples $(C, V, c)$ for which this problem is particularly easy to solve.

For any $d \in C - \{c\}$, let Deficit$[d]$ be the number of votes $c$ needs to gain in order to have more votes than $d$ in a pairwise contest between $c$ and $d$.

**Definition A.2** *Any Dodgson triple $(C, V, c)$ is said to be* nice *if for each candidate $d \in C - \{c\}$, there are at least* Deficit$[d]$ *votes for which candidate $c$ is exactly one position below candidate $d$.*

Given a Dodgson triple $(C, V, c)$, the algorithm Greedy-Score works as follows:

1. For each candidate $d \in C - \{c\}$, determine Deficit$[d]$.

2. If $(C, V, c)$ is not nice then output ("anything","maybe"); otherwise, output

$$(\textstyle\sum_{d \in C - \{c\}} \text{Deficit}[d], \text{"definitely"}).$$

Note that, for nice Dodgson triples, we have

$$\text{DodgsonScore}(C, V, c) = \sum_{d \in C - \{c\}} \text{Deficit}[d],$$

It is easy to see that Greedy-Score is a self-knowingly correct polynomial-time bounded algorithm. To show that it is even *frequently* self-knowingly correct, Homan and Hemaspaandra prove the following lemma. Their proof uses a variant of Chernoff bounds.

**Lemma A.3 (see Thm. 4.1(3) in [HH06])** *Let $(C, V, c)$ be a given Dodgson triple with $n = \|V\|$ votes and $m = \|C\|$ candidates, chosen uniformly at random among all such Dodgson elections. The probability that $(C, V, c)$ is not nice is at most $2(m-1)e^{-\frac{n}{8m^2}}$.*

Homan and Hemaspaandra [HH06] show that the heuristic Greedy-Winner, which is based on Greedy-Score and which solves the winner problem for Dodgson elections, also is a frequently self-knowingly correct polynomial-time algorithm. This result is stated formally below.

**Theorem A.4 ([HH06])** *For all $m, n \in \mathbb{N}^+$, the probability that a Dodgson election $(C, V)$ selected uniformly at random from all Dodgson elections having m candidates and n votes (i.e., all $(m!)^n$ Dodgson elections having m candidates and n votes have the same likelihood of being selected) has the property that there exists at least one candidate c such that Greedy-Winner on input $(C, V, c)$ outputs "maybe" as its second output component is less than $2(m^2 - m)e^{-\frac{n}{8m^2}}$.*

# B  Foundations of Average-Case Complexity Theory

The theory of average-case complexity was initiated by Levin [Lev86]. A problem's average-case complexity can be viewed as a more significant measure than its worst-case complexity in many cases, for example in cryptographic applications. We here follow Goldreich's presentation [Gol97]. Another excellent introduction to this theory is that of Wang [Wan97].

Fix the alphabet $\Sigma = \{0, 1\}$, let $\Sigma^*$ denote the set of strings over $\Sigma$, and let $\Sigma^n$ denote the set of all length $n$ strings in $\Sigma^*$. For any $x, y \in \Sigma^*$, $x < y$ means that $x$ precedes $y$ in lexicographic order, and $x - 1$ denotes the lexicographic predecessor of $x$.

Intuitively, Levin observed that many hard problems—including those that are NP-hard in the traditional worst-case complexity model—might nonetheless be easy to solve "on the average," i.e., for "most" inputs or for "most practically relevant" inputs. He proposed to define the complexity of problems with respect to some suitable distribution on the input strings.

We now define the notion of a distributional problem and the complexity class AvgP.

Here, we define only distributional search problems; the definition of distributional decision problems is analogous.

**Definition B.1 ([Lev86], see also [Gol97, Wan97])**   *1. A* distribution function *$\mu$ : $\Sigma^* \to [0, 1]$ is a nondecreasing function from strings to the unit interval that converges to one (i.e., $\mu(0) \geq 0$, $\mu(x) \leq \mu(y)$ for each $x < y$, and $\lim_{x \to \infty} \mu(x) = 1$). The* density function *associated with $\mu$ is defined by $\mu'(0) = \mu(0)$ and $\mu'(x) = \mu(x) - \mu(x - 1)$ for each $x > 0$. That is, each string $x$ gets weight $\mu'(x)$ with this distribution.*

2. A distributional (search) problem *is a pair* $(f, \mu)$, *where* $f : \Sigma^* \to \Sigma^*$ *is a function and* $\mu : \Sigma^* \to [0, 1]$ *is a distribution function.*

3. A function $t : \Sigma^* \to \mathbb{N}$ *is* polynomial on the average *with respect to some distribution* $\mu$ *if there exists a constant* $\epsilon > 0$ *such that*

$$\sum_{x \in \Sigma^*} \mu'(x) \cdot \frac{t(x)^\epsilon}{|x|} < \infty.$$

4. *The class* AvgP *consists of all distributional problems* $(f, \mu)$ *for which there exists an algorithm* $\mathcal{A}$ *computing* $f$ *such that the running time of* $\mathcal{A}$ *is polynomial on the average with respect to the distribution* $\mu$.

In Section 3.2, we focused on the standard uniform distribution $\hat{\mu}$ on $\Sigma^*$, which is defined by

$$\hat{\mu}'(x) = \frac{1}{|x|(|x|+1)2^{|x|}}.$$

That is, we first choose an input size $n$ at random with probability $1/(n(n + 1))$, and then we choose an input string of that size $n$ uniformly at random. (In this model, the length-0 string $\epsilon$ is routinely completely excluded from the probability distribution—it is by convention given weight zero.) For each $n \in \mathbb{N}^+$, let $\hat{\mu}_n$ be the restriction of $\hat{\mu}$ to strings of length exactly $n$. For each $n \in \mathbb{N}^+$, let $\hat{\mu}_{\leq n}$ be the restriction of $\hat{\mu}$ to strings of length at most $n$.

In Section 3.2, we considered polynomial-time benign algorithm schemes. This notion was introduced by Impagliazzo [Imp95] to provide an alternative view on the definition of average polynomial time.

**Definition B.2 ([Imp95])**  1. *An algorithm computes a function* $f$ *with* benign faults *if it either outputs an element of the image of* $f$ *or "?," and if it outputs anything other than ?, it is correct.*

2. A polynomial-time benign algorithm scheme *for a function* $f$ *on* $\mu_n$ *is an algorithm* $\mathcal{A}(x, \delta)$ *such that:*

   (a) $\mathcal{A}$ *runs in time polynomial in* $|x|$ *and* $1/\delta$.
   
   (b) $\mathcal{A}$ *computes* $f$ *with benign faults.*
   
   (c) *For each* $\delta$, $0 < \delta < 1$, *and for each* $n \in \mathbb{N}^+$,
   
   $$\mathrm{Prob}_{\hat{\mu}_{\leq n}}[\mathcal{A}(x, \delta) = ?] \leq \delta.$$